\documentclass[pdflatex,sn-nature]{sn-jnl}
\catcode`\|=12\relax
\usepackage{anyfontsize}
\usepackage{siunitx}
\usepackage{graphicx}
\usepackage{multirow}
\usepackage{amsmath,amssymb,amsfonts}
\usepackage{amsthm}
\usepackage{mathrsfs}
\usepackage[title]{appendix}
\usepackage{xcolor}
\usepackage{textcomp}
\usepackage{manyfoot}
\usepackage{booktabs}
\usepackage{algorithm}
\usepackage{algorithmicx}
\usepackage{algpseudocode}
\usepackage{listings}
\usepackage{graphicx}
\usepackage{url}
\usepackage{longtable}
\usepackage{tabularx}
\usepackage{listings}
\usepackage{xcolor} 
\usepackage{colortbl}
\usepackage{bibunits}
\makeatletter
\newcounter{savebibitem}
\let\oldthebibliography\thebibliography
\renewcommand{\thebibliography}[1]{%
  \oldthebibliography{#1}%
  \ifnum\value{savebibitem}>0
    \setcounter{NAT@ctr}{\value{savebibitem}}%
  \fi
}
\let\oldendthebibliography\endthebibliography
\renewcommand{\endthebibliography}{%
  \setcounter{savebibitem}{\value{NAT@ctr}}%
  \oldendthebibliography%
}
\makeatother
\theoremstyle{thmstyleone}%

\theoremstyle{thmstyletwo}%

\theoremstyle{thmstylethree}%

\usepackage{geometry}
\begin{document}

\title[Sycophantic AI makes human interaction feel more effortful and less satisfying over time]{\textsc{Sycophantic AI makes human interaction feel more effortful and less satisfying over time}}

\author*[1]{\fnm{Lujain} \sur{Ibrahim}}\email{lujain.ibrahim@oii.ox.ac.uk}

\author[1]{\fnm{Franziska Sofia} \sur{Hafner}}

\author[2]{\fnm{Myra} \sur{Cheng}}
\author[2]{\fnm{Cinoo} \sur{Lee}}
\author[1,3]{\fnm{Rebecca} \sur{Anselmetti}}
\author[2]{\fnm{Robb} \sur{Willer}}

\author[1]{\fnm{Luc} \sur{Rocher}}

\author[2]{\fnm{Diyi} \sur{Yang}}\email{diyiy@stanford.edu}

\affil[1]{\orgname{University of Oxford}}

\affil[2]{\orgname{Stanford University}}

\affil[3]{\orgname{UK AI Security Institute}}

\begin{bibunit}[naturemag]

\abstract{
Millions of people now turn to artificial intelligence (AI) systems for personal advice, guidance, and support. Such systems can be sycophantic, frequently affirming users' views and beliefs. Across five preregistered studies (N = 3,075 participants, 12,766 human-AI conversations), including a three-week study with a census-representative U.S. sample, we provide longitudinal experimental evidence that sycophantic AI shifts how users approach their closest relationships. We show that sycophantic AI immediately delivers the emotional and esteem support users typically associate with close friends and family. Over three weeks of such interactions, users became nearly as likely to seek personal advice from sycophantic AI as from close friends and family, and reported lower satisfaction with their real-world social interactions. When given a choice among AI response styles, a majority preferred sycophantic AI---not for the quality of its advice, but because it made them feel most understood. Together, these findings offer a relational account of AI sycophancy and its impacts.
}

\maketitle

\section*{Introduction}\label{sec1}

When facing a difficult job decision, a struggling relationship, or an everyday worry, people have long turned to those close to them for comfort and advice. Today, AI systems are rapidly taking on this role. Already, 12\% of U.S. teens and over 20\% of young adults turn to AI chatbots for personal support, and many young people report finding it easier to discuss mental health with an AI chatbot than with a doctor or psychologist~\cite{mcclain2026teens,Zao-Sanders_2025,mcbain2025genai, ipsos2026jeunesia}. A growing body of evidence shows that, in such conversations, AI systems tend to be sycophantic, frequently affirming users’ views and beliefs~\cite{cheng2026elephant,sharma2023towards}. Even a single conversation with sycophantic AI has been shown to influence users’ judgment, leading them to hold more ideologically extreme political views~\cite{rathje2025sycophantic}, to be less willing to consider their own role in conflicts~\cite{cheng2026sycophantic} and, in severe cases, to be more likely to engage in self-harm~\cite{hill2025teen}. 

Previous work has studied such effects resulting from single conversations, but in reality, people confide in AI systems repeatedly over weeks and months, just as they do with the people in their lives. At this scale, the actual influence of sycophantic AI depends not on any isolated conversation, but on how that influence unfolds alongside users' relationships with friends, family, and partners. Even if sycophantic AI distorts a user's judgment, these human relationships can still offer honest reflection, pushback, and perspective, buffering against its effects. At the same time, repeated exposure to unconditional AI affirmation may itself alter how users experience conversations with the people around them, shifting what feels sufficiently understanding, satisfying, or worth seeking out~\cite{perry2026defense}. Thus, the central long-term risk of sycophantic AI may not be the judgments it distorts in any single conversation, but how it gradually reshapes the very relationships that would otherwise constrain its influence.

Here, in five pre-registered studies ($N = 3,075$ unique participants, $N=12,766$ human-AI conversations), including a three-week longitudinal study with a census-representative U.S. sample, we investigate how repeated discussion of personal dilemmas with sycophantic AI shapes users' close relationships. We first establish that there is a gap between what people say they want from AI versus close friends and family, and what sycophantic AI actually provides. Participants reported seeking distinct kinds of support from each, yet in practice, sycophantic AI reliably delivered the forms of support, such as feeling understood and validated, that people more strongly associate with close human relationships. In turn, interactions with sycophantic AI shifted how participants approached these close relationships. Having just experienced active affirmation, they anticipated needing more effort to feel understood by the people closest to them, and felt less need to talk through their dilemmas any further.

Tracking these effects longitudinally over weeks of use, we find that while sycophantic AI consistently made participants feel good in the moment, it produced none of the downstream benefits that such support from humans typically provides~\cite{reis2004perceived,reis2018perceived}: we observed no increases in intellectual humility or connection to real-world relationships. Instead, these participants reported lower satisfaction with their real-world social interactions over the course of the study. Even so, by the study's end, they were nearly as likely to seek personal advice from sycophantic AI as from their close friends and family.

We further find that even when AI does not default to sycophancy, users may actively seek it out: when given a choice, a majority chose sycophantic AI over more balanced alternatives, not for the quality of its advice, but for how easy the conversation was and how understood it made them feel. Taken together, our findings offer a \textit{relational} account of AI sycophancy and its impact: by providing effortless understanding, sycophantic AI can raise the bar against which human relationships are judged. Over time, if sycophantic AI becomes a comparable alternative to close friends and family, it could reshape patterns of social support at scale~\cite{ong2026friendlier}. As the AI industry shifts toward greater personalization and user choice, our findings underscore the importance of both user-side and model-level mitigations of AI sycophancy.

\section{Defining and operationalizing sycophancy}
Sycophancy has been broadly defined as AI systems agreeing with users even when that agreement is not warranted. However, it has been operationalized very differently across domains: from agreeing with factually incorrect claims~\cite{sharma2023towards}, to endorsing users' actions in personal conflicts~\cite{cheng2026sycophantic}, to supporting users' preexisting political viewpoints~\cite{rathje2025sycophantic}, to failing to push back on users' framing of situations in advice-seeking queries~\cite{cheng2026elephant}. Here, we operationalize sycophancy as active affirmation of user views and reasoning. As active affirmation and active challenge may each have their own effects on users, and our primary interest is in isolating the effects of active affirmation specifically, we compare a \textit{sycophantic} AI, designed to agree with and support a user's views, and a \textit{neutral} AI, designed to remain impartial and present multiple perspectives without explicitly agreeing or disagreeing. In Studies 4 and 5, we additionally include a \textit{challenging} AI, designed to question users' views and offer counterarguments, to provide a more comprehensive view of how different engagement styles compare. All three conditions shared identical instructions on tone, formatting, and conversational style, differing only in how they responded to participants' views (see Supplementary Methods, Sections 1.1 for full prompts). We validated that each model engaged as intended through both computational analyses (i.e., measuring the prevalence of specific sycophantic behaviors in model responses) and participant perceptions (see Supplementary Information, Section 2.1). 

\section{Results}
\begin{figure}[!htbp]
\centering
\includegraphics[width=1\linewidth]{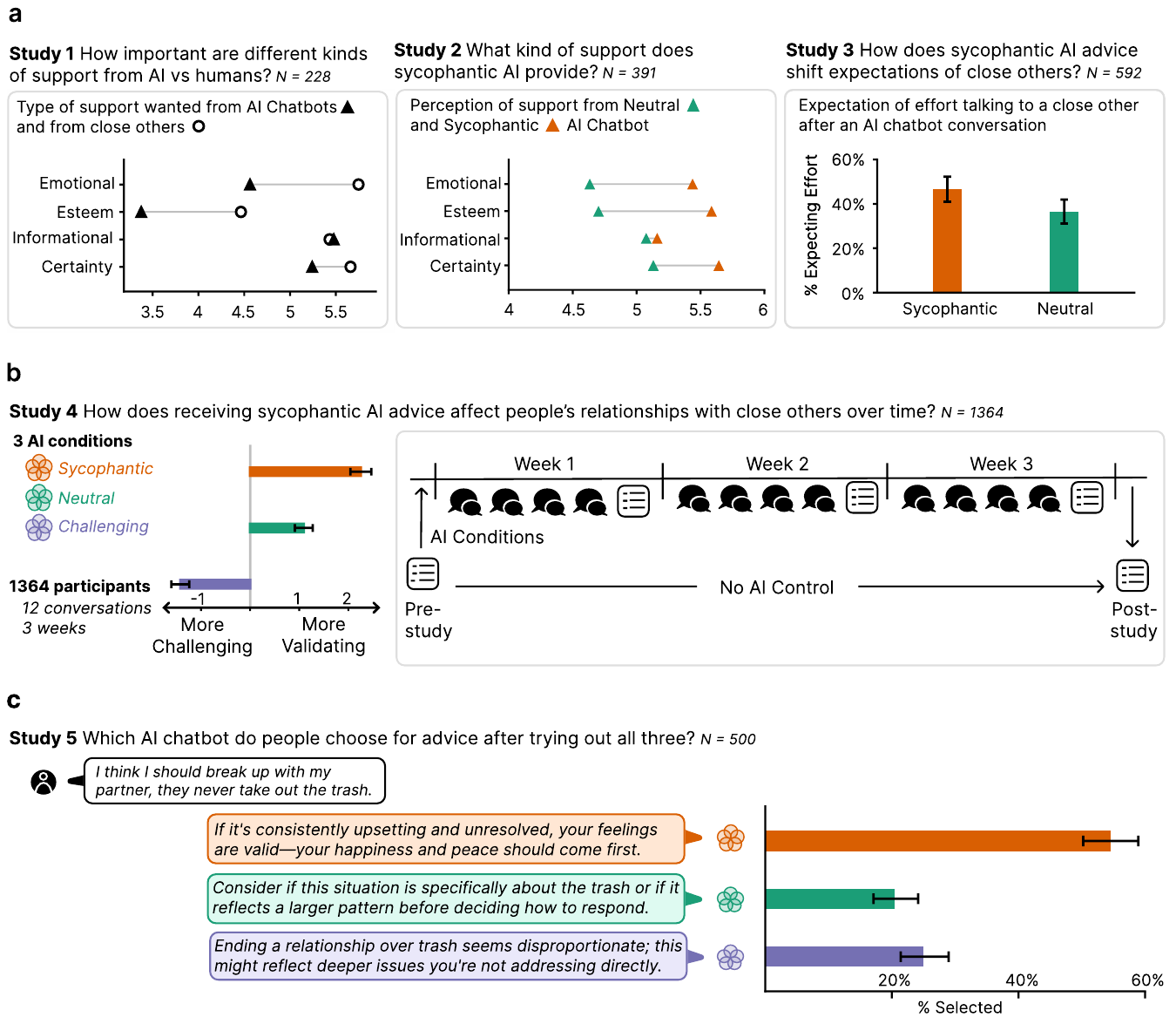}
\caption{\textbf{Overview of five preregistered studies on the impact of sycophantic AI on human relationships.} (A) Studies 1–3: Expectations of support from humans vs. AI systems. In all studies, participants first describe a personal situation they are seeking advice on. In Study 1, they are then randomly assigned to rate how much they would expect each type of support (e.g., informational, emotional) from either an AI system or a close other. In Study 2, they are randomly assigned to discuss the situation with either a sycophantic or neutral AI and then rate the AI systems on the same support dimensions. In Study 3, participants identify a close other they would normally go to for advice on such a situation, are randomly assigned to discuss the situation with either a sycophantic or neutral AI, and are then asked to answer questions about how they would expect the conversation with their identified close other to go. (B) Study 4: Longitudinal exposure to sycophantic AI advice. Participants are randomly assigned to one of three AI conditions (sycophantic, neutral, or challenging) or a no-AI control group. All participants complete a pre-study survey at the beginning of the experiment. Participants in the AI conditions then converse with their assigned AI system every other day over three weeks, completing surveys at the end of each session, each week, and the entire study period. At the end of the study, participants rated their assigned AI on validation (explicit and implicit) and challenge in a manipulation check. The average rating of validation minus the average rating of challenge is plotted in the bar plot for each of the sycophantic, challenging, and neutral AI. (C) Study 5: User choice dynamics across different interaction styles. Participants converse with unlabeled neutral, sycophantic, and challenging AIs in randomized order, then choose which AI system they would most want to continue talking to. Sample sizes for each study were determined using power analyses, with details in the pre-registrations and Supplementary Information.}
\label{fig:experiments_overview}
\end{figure}

We conducted five preregistered studies ($N=3,075$; see Figure~\ref{fig:experiments_overview}) to investigate the relational consequences of sycophantic AI. Studies 1 and 2 first establish what kinds of support people associate with close others vs. AI, and what support people perceive sycophantic AI to provide. Study 3 tests whether a single interaction with sycophantic AI shifts how people anticipate subsequent conversations with close others to go. Importantly, capturing sycophancy's downstream effects requires a longitudinal design. Thus, Study 4 tests whether sycophancy's effects accumulate into measurable changes in users' attitudes towards their relationships over time. Finally, given that users now have growing control over AI behavior, Study 5 asks whether, despite these relational consequences, users nonetheless choose to seek advice from sycophantic AI over other styles.

\subsection*{Sycophantic AI provides the support people associate with close human relationships}
\begin{figure}[!htbp]
\centering
\includegraphics[width=1\linewidth]{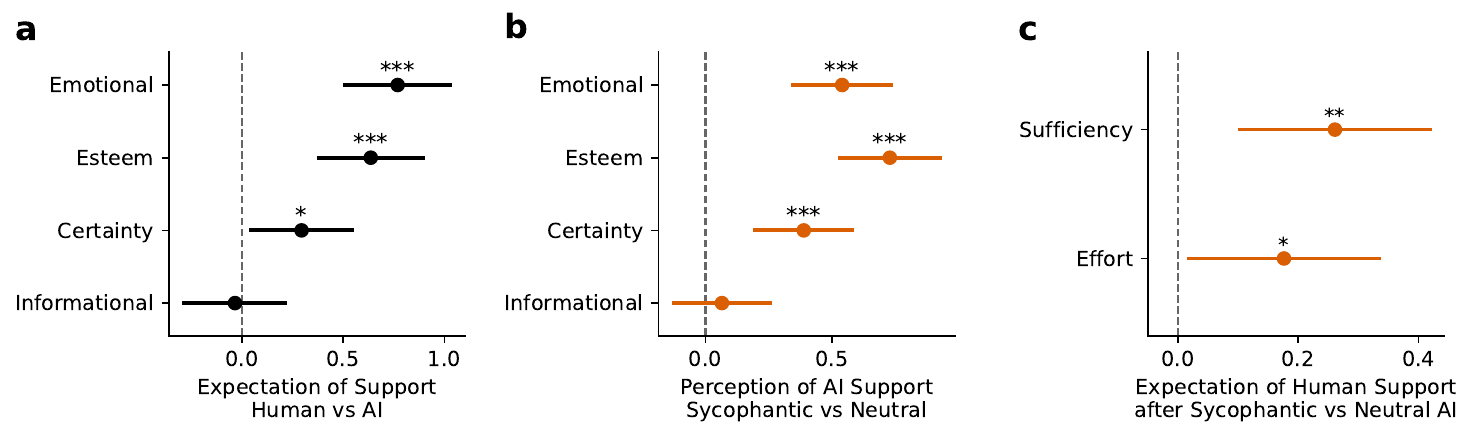}
\caption{\textbf{Sycophantic AI provides the types of support people associate with close human relationships.} All studies used between-subjects random assignment. (A) In Study 1, participants ($N=228$) rated how important receiving different types of support would be when discussing a personal situation either with close others or an AI system. Participants valued emotional support (understanding and care), esteem support (validation and recognition of strengths), and certainty (confidence and clarity on next steps) more from humans than from AI systems, with no statistically significant difference for informational support (practical advice and new perspectives). (B) In Study 2, participants ($N = 391$) discussed a personal situation with a sycophantic or neutral AI, then rated the AI systems on each type of support. Participants rated the sycophantic AI (vs. neutral) higher on emotional support, esteem support, and certainty, with no statistically significant difference for informational support. (C) In Study 3, participants ($N = 592$) discussed a personal situation with a sycophantic or neutral AI and then reported expectations for a subsequent conversation with a chosen confidant. Participants in the sycophantic (vs. neutral) condition anticipated that greater effort would be needed to be understood by this confidant and reported greater conversational sufficiency, feeling they had already talked through their situation enough. Effect sizes are Cohen's $d$ with 95\% CIs (*$p<0.05$, **$p<0.01$, ***$p<0.001$)}.
\label{fig:combined_effects_study_1_2_3}
\end{figure}

Study 1 first establishes whether people turn to AI systems and close others for different kinds of support when seeking advice. We asked participants ($N = 228$) to describe a personal situation they would like advice on and then rate the importance of receiving four types of support, either from an identified close person in their life or from an AI system: \textit{emotional} support (understanding and care), \textit{esteem} support (validation and recognition of strengths), \textit{informational} support (practical advice and new perspectives), and \textit{certainty} (confidence and clarity on next steps)~\cite{cheng2026verbalizing,cutrona1992controllability}. Figure~\ref{fig:combined_effects_study_1_2_3} shows that desired support from AI systems and close others varied significantly by support type ($F(3,678) = 17.69$, $p <0.001$, partial $\eta^2$ = 0.073). People valued far more emotional ($d = 0.77$, $p < 0.001$) and esteem support ($d = 0.64$, $p < 0.001$) from humans than from AI systems, valued certainty somewhat more from humans ($d = 0.29$, $p = 0.027$), and showed no statistically significant difference for informational support ($d = -0.04$, $p = 0.789$). Thus, people approach AI systems and close others for distinct kinds of support, looking to close others especially for emotional and esteem support.

Given this gap, Study 2 asks whether sycophantic AI may close it, providing, in practice, the kinds of support people associate with close human relationships. Participants ($N = 391$) described a personal situation they would like advice on, discussed it with either a sycophantic or neutral AI, and then rated the AI systems on the same four types of support from Study 1. Participants in the sycophantic (compared to neutral) condition rated the AI higher on emotional support ($d = 0.54$, $p < 0.001$), esteem support ($d = 0.73$, $p < 0.001$), and certainty ($d = 0.39$, $p < 0.001$), with no statistically significant difference in informational support ($d = 0.07$, $p = 0.5$)~\cite{yin2024ai}. Sycophantic AI thus delivers what we will call \textit{relational} support---emotional and esteem support, the two types people most strongly associate with close others rather than AI systems.

\subsection*{After sycophantic AI, people expect more effort to feel understood by humans}
In Study 3, we ask whether receiving this relational support from sycophantic AI shapes how people approach subsequent conversations with close others. Participants ($N = 592$) described a personal situation they would like advice on, discussed it in a live interaction with either a sycophantic or neutral AI, and then rated their expectations for a subsequent conversation about the same situation with a chosen confidant (i.e., a friend, partner, or family member). Participants in the sycophantic (vs. neutral) condition anticipated that it would take greater effort to be understood by this person ($d = 0.18$, $p = 0.03$). They also reported greater conversational sufficiency following the sycophantic AI conversation, feeling they had already talked through their personal situation enough ($d = 0.26$, $p = 0.002$). These effects were more pronounced when the confidant identified was a friend or family member rather than a romantic partner (effort $d = 0.21$; sufficiency $d = 0.29$), suggesting that sycophantic AI may compete differently with different types of human relationships.

\subsection*{Across weeks, participants were nearly as inclined to seek advice from sycophantic AI as from close others}
Study 4 tests whether sustained use of sycophantic AI impacts advice-seeking preferences and social outcomes over time. For three weeks, participants ($N = 1364$) discussed personal advice topics such as relationships and everyday habits with a sycophantic, neutral, or challenging AI, or were placed in a no-AI control condition (full list of 16 topic choices in Supplementary Methods, Section 1.3). Figure~\ref{fig:combined_figure_longitudinal} shows that compared to the neutral condition, participants in the sycophantic condition showed a smaller gap between their inclination to seek advice from AI and from close friends and family (0.37 pt on a 7-point scale; $d = 0.19$, $p_{\text{adj}} = 0.034$). This shift emerged early and persisted across all three weeks. It operated through relational rather than epistemic pathways. The preference for AI vs. close others was mediated by feeling understood by the AI (indirect $= 0.22$, 95\% CI [0.06, 0.38], 59\% mediated), perceived helpfulness of the conversations (indirect $= 0.18$, 95\% CI [0.04, 0.31], 48\% mediated), and positive affect (indirect $= 0.13$, 95\% CI [0.01, 0.26], 34\% mediated), rather than by increased certainty.

\begin{figure}[!htbp]
\centering
\includegraphics[width=0.95\linewidth]{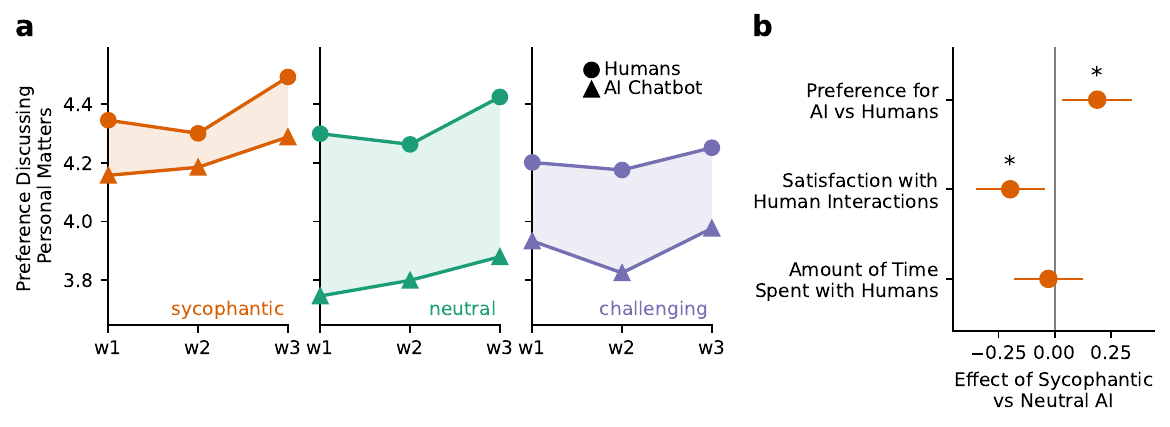}
\caption{\textbf{Sycophantic AI narrows the gap with close others as advice sources and reduces real-world social satisfaction.} In a three-week longitudinal study, participants ($N = 1364$) were randomized to interact with a sycophantic, neutral, or challenging AI, or assigned to a no-AI control condition, and discussed personal topics (e.g., relationship advice, career decisions, personal habits). (A) Weekly trajectories of participants' inclination to turn to their assigned AI system vs. close others for personal advice. The sycophantic condition showed a smaller gap between AI and humans than the neutral condition. (B) Compared to the neutral condition, participants in the sycophantic condition reported lower satisfaction with their real-world social interactions over the three weeks, and a greater shift in advice-seeking toward their assigned AI system, with no statistically significant effect on the social time they spent with others. Effect sizes are Cohen's d with 95\% CIs (*$p <0.05$, **$p < 0.01$, ***$p<0.001$).}
\label{fig:combined_figure_longitudinal}
\end{figure}

\begin{figure}[!htbp]
\centering
\includegraphics[width=1\linewidth]{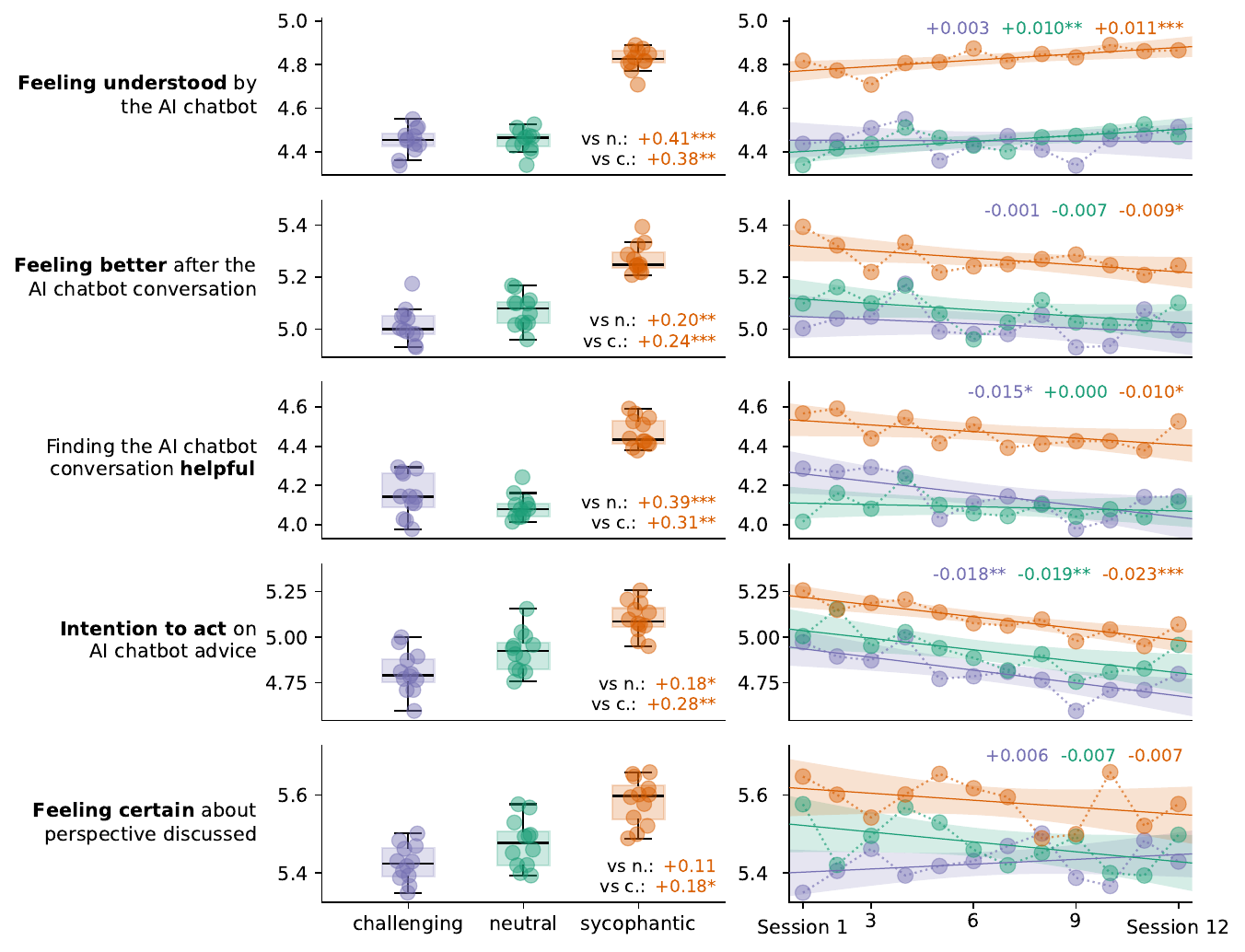}

\caption{\textbf{Sycophantic AI is consistently perceived as more understanding and helpful across repeated interactions.} Session-level outcomes from the longitudinal study ($N = 1364$), where participants had one conversation per session with their assigned AI system. Each row shows one session-level outcome, with the left column showing overall means by condition and the right column showing trajectories across sessions. Participants in the sycophantic (vs. neutral and challenging) condition consistently reported feeling more understood by their assigned AI chatbot. They also found the conversations more helpful, felt better after having them, and reported stronger intentions to act on the chatbot's advice. Feeling understood was the only measure that increased over time. While participants in the sycophantic condition felt more certain in their perspectives than those in the challenging condition, they did not differ significantly from those in the neutral condition. In the left column, center lines indicate medians, boxes indicate interquartile ranges, and whiskers extend to the most extreme data point within 1.5x of the interquartile range. In the right column, error bars and shaded regions represent 95\% confidence intervals (*$p <0.05$, **$p < 0.01$, ***$p<0.001$).}
\label{fig:session_level_overview}
\end{figure}

\subsection*{Feeling understood by AI grows with sustained use but does not provide downstream benefits}
Figure~\ref{fig:session_level_overview} shows that participants in the sycophantic and neutral conditions reported small but statistically significant increases in feeling understood by AI over the three-week period, with those in the sycophantic condition showing a higher baseline and a steeper trajectory. This was the case even though the chat history was reset after each conversation (i.e., the AI systems retained no ``memories" of previous conversations). Other session-level measures, such as affect or certainty, remained flat or slightly declined over the three weeks. In human relationships, feeling understood and supported is broadly associated with downstream benefits for how people see themselves, their relationships, and the world~\cite{reis2018perceived,reis2004perceived, yang2016social, smith2008social, li2026random}. Here, we find that feeling understood by sycophantic AI was largely contained to the AI interaction itself. Compared to the neutral condition, the sycophantic condition showed no statistically significant effect on feeling understood by other humans ($d = -0.08$, $p = 0.295$) or intellectual humility ($d = 0.01$, $p = 0.861$). We also tested whether sycophantic AI inflated participants' self-perceptions---a possible consequence of receiving validation, and the opposite of intellectual humility~\cite{rathje2025sycophantic}. Sycophantic AI did not lead participants to rate themselves more favorably than the average person ($d = 0.06$, $p = 0.263$), but did show a marginal tendency for participants to rate themselves more favorably than the specific people in their lives ($d = 0.11$, $p = 0.080$).

\subsection*{Sycophantic AI reduces satisfaction with social interactions}
Participants in the sycophantic condition reported lower satisfaction with their real-world social interactions (a preregistered exploratory measure; 5.51 vs.\ 5.70 on a 7-point scale; $d = 0.20$, $p_\text{adj} = 0.022$) compared to those in the neutral condition. This effect was not accompanied by social withdrawal or reduced time spent with others ($d = -0.03$, $p_\text{adj} = 0.719$), suggesting that the shift was perceptual rather than behavioral. Sycophancy narrowed the gap between how understood participants felt by AI vs. humans, from 0.56 points in the neutral condition to just 0.13 points in the sycophantic condition ($d = 0.25$, $p_\text{adj} < 0.001$),  meaning that participants in the sycophantic condition felt nearly as understood by their AI chatbot as by the people in their lives. The shift in social satisfaction was mediated by this narrowing gap, not by feeling understood by AI on its own (indirect effect $= -0.079$, 95\% CI $[-0.135, -0.029]$, 41\% mediated).\footnote{A possible explanation for the decrease in social satisfaction is that participants who discussed interpersonal conflict with a sycophantic AI may have come away with more negative views of the people involved. In an additional, preregistered study ($N=1,099$, see Supplementary Information, Section 2.7), we tested this directly and found that sycophantic AI (compared to neutral) did not increase attributions of blame towards the other person or reduce perspective-taking in conflict scenarios after a single conversation.} After controlling for the gap, the direct effect of sycophancy on social satisfaction was no longer significant ($p = 0.13$), suggesting that what reduced satisfaction was not how the AI felt on its own, but how close it came to the humans in participants' lives.
\begin{figure}[!htbp]
\centering
\includegraphics[width=1\linewidth]{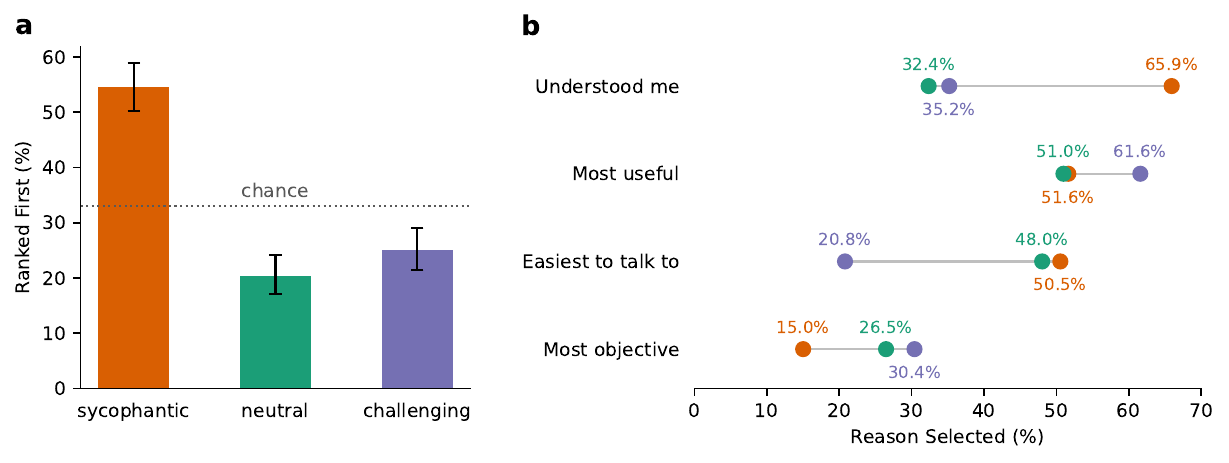}
\caption{\textbf{When given a choice, users select sycophantic AI over neutral and challenging alternatives.} Participants ($N = 500$) had short conversations with three AI systems (sycophantic, neutral, and challenging, unlabeled and in randomized order) and selected which one they would most want to continue talking to. (A) Proportion of participants selecting each AI system. A majority ($54.6\%$) chose the sycophantic AI, significantly above chance. (B) Reasons participants gave for their choice, by chosen AI system. Participants who chose the sycophantic AI were more likely to report that it \textit{understood me best} and \textit{was easiest to talk to}, while those who chose the challenging AI were more likely to report that it \textit{gave the most objective advice}. Error bars represent 95\% confidence intervals.}
\label{fig:combined_first_and_reasons}
\end{figure}

\subsection*{Users choose to interact with sycophantic AI even after trying alternatives}
Studies 1-4 establish the relational consequences of sycophantic AI. In Study 5, we test whether users prefer sycophantic AI when they directly interact with multiple AI systems with different styles and make an active choice between them. Participants ($N = 500$) had short conversations with three AI systems (sycophantic, neutral, and challenging) then selected the one they would most want to continue talking to. A majority ($54.6\%$) chose the sycophantic AI, significantly above chance ($p < 0.001$; $\chi^2(2) = 103.35$, $w = 0.46$). This preference was robust across the topics participants chose to discuss ($\chi^2(20) = 26.90$, $p = 0.14$) and the order in which they sampled the three AI systems ($p = 0.81$, two-sided binomial test that the chosen AI system is the first-sampled one). Figure~\ref{fig:combined_first_and_reasons} shows that participants who chose the sycophantic AI disproportionately reported that it \textit{understood me best} ($\chi^2(1) = 32.77$, $p < 0.001$ vs.\ neutral; $\chi^2(1) = 31.68$, $p < 0.001$ vs.\ challenging) and was \textit{easiest to talk to} ($\chi^2(1) = 30.11$, $p < 0.001$ vs.\ challenging), rather than that \textit{it gave the most useful advice}, which did not differ across choice groups ($\chi^2(2) = 3.89$, $p = 0.14$). In contrast, those who chose the challenging AI were more likely than those who chose the sycophantic AI to report that it \textit{gave the most objective advice} ($\chi^2(1) = 11.80$, $p < 0.001$).

\section*{Discussion}
Prior work has focused on measuring sycophancy in LLMs, understanding why it emerges, and assessing its influence on users in single interactions~\cite{ibrahim2025training, sharma2023towards, rathje2025sycophantic, cheng2026sycophantic, dubois2026ask, cheng2026verbalizing, shen2026guidance, sun2025friendly, bo2026invisible}. Here, we move beyond isolated interactions to provide the first longitudinal experimental evidence of sycophantic AI's effects on users. We also compare sycophantic AI to a neutral AI rather than to a default or challenging baseline, isolating the effect of active affirmation from the absence of challenge. Focusing on the personal advice domain, we show that interactions with sycophantic AI can lead people to anticipate greater effort to be understood in their closest relationships, leave them less satisfied with their real-world social interactions, and over time, shift them toward treating sycophantic AI as a source of advice comparable to friends and family.\footnote{See Supplementary Information, Section 2.5.12 for a content analysis of the advice participants received across different AI conditions. Advice from the sycophantic AI was less prosocial and more self-focused than advice from the other conditions.} Taken together, our findings advance a \textit{relational} account of AI sycophancy and its impacts, with important implications for the development, alignment, and governance of AI systems used for personal guidance.

Recent analyses of tens of thousands of real-world human-AI conversations found that relationship advice is the domain where sycophantic AI behavior occurs most often~\cite{shen2026guidance}. While public concerns around AI sycophancy have focused on high-profile cases with severe outcomes, such as the reinforcement of delusions and suicidal ideation~\cite{Yang_2024,hill2025teen,moore2026characterizing}, our findings demonstrate that sycophancy can have lasting effects even on the median user.\footnote{Across studies, pre-registered moderator analyses showed that the effects of sycophantic AI did not concentrate among participants with weaker social ties, more agreeable personalities, closer relationships with their selected confidants, or higher prior trust in AI. In Study 2, neither AI-usage frequency nor relationship type with the confidant selected moderated participants' ratings. In Study 3, baseline interpersonal closeness did not moderate either primary outcome. In Study 4, social network strength, agreeableness, and prior trust in AI advice did not moderate any primary outcomes.} This expands the policy relevance of sycophancy from rare safety failures to possible population-level social effects, where even small individual-level effects like those we observe can accumulate into meaningful consequences across large user bases of hundreds of millions. It also highlights the importance of further investigating for whom, under what conditions, and through what mechanisms sycophantic AI shapes social outcomes. The consequences of using sycophantic AI may also extend beyond how one receives support to how one gives it~\cite{perry2023ai}: if exposure to sycophancy reduces the frequency or depth of conversations people have with others, it may shape perspective-taking and other social skills. Our sample also reflects a particular cultural context, and future work should examine populations with different norms around social support, where effects could differ in direction or magnitude~\cite{taylor2004culture, bakir2024manipulation}.

The effects we observed emerged after only three weeks of exposure, meaning whether they intensify or plateau over months or years of AI use remains an open question. Consistent with some recent longitudinal studies of a similar timescale, we find that the effect of sycophantic AI did not differ from that of neutral or challenging AI on affective well-being~\cite{kirk2025neural, luettgau2025people, phang2025investigating}. Recent surveys have linked having frequent personal conversations with AI to increased anxiety and depression~\cite{perlis2026generative, zhang2025rise}, though the directionality of these associations remains unclear. Assessing this class of risks will require evaluation paradigms built for sustained interaction, not single-turn tests~\cite{ibrahim2025towards,wei2026cascades, hwang2025ai}. Given that personal use of AI systems will likely continue to grow---and may offer genuine long-term benefits if designed well---building AI systems that support rather than undermine users' broader well-being will require further investment in longitudinal experimental designs and new methods for measuring effects as they unfold over time. Such efforts will benefit from regular input from clinicians, social scientists, and other domain experts who have insight into possible social and psychological consequences~\cite{zhao2026psychologists}.

Beyond the question of how these effects evolve over time, the relational dynamics our work identifies could intensify as AI capabilities and affordances advance. AI systems are increasingly equipped with personalization and persistent memory, recalling users' preferences, past decisions, and emotional states across conversations~\cite{wu2025human,karny2026neural}.  Recent evidence suggests personalization may directly amplify sycophancy~\cite{jain2026interaction}. However, it may also produce similar relational effects on its own. Personalized AI systems will be able to remember everything, require little self-disclosure, and integrate digital traces such as emails or documents to model users in great detail~\cite{gabriel2025we, shaikh2025creating}. Thus, they may be similarly able to provide the experience of being seen and understood without much effort or friction, regardless of whether overt sycophancy is present. Beyond personalization, as AI systems take on embodied and multimodal forms, the societal impact of these affordances will warrant similar investigation.

Our findings also point toward possible mitigations. We show that giving users a choice between neutral, challenging, and sycophantic AI systems does not reduce preferences for sycophancy. Combined with evidence that users can actively push AI systems toward validation in real-world conversations~\cite{shen2026guidance}, this suggests that user-side mitigations alone, such as offering AI personality or style options, are unlikely to be sufficient. Future work should test whether transparency about interaction style, reflective prompts, or warnings about downstream consequences can shift user preferences, but preventing these dynamics will likely depend primarily on model-side mitigations~\cite{perry2026defense, zohar2026against}.

Sycophantic AI delivers what people have always sought from close others---the experience of being seen and understood---but without the work that produces it: the time to explain and listen, the risk of opening up, and the empathic effort of bridging disparate experiences. When feeling understood becomes the default of every interaction, the human relationships that still require the work may, over time, come to feel like insufficient versions of what AI systems readily provide. As AI takes on a growing role in everyday life, whether it strengthens or damages human relationships will depend on designing these systems to help users without quietly leading them to expect from one another what only AI can deliver.
\section*{Methods}

\subsection*{AI manipulation}
Across all studies, we varied AI model behavior using system prompts with the same underlying model (\texttt{gpt-4o-2024-11-20}). We set generation parameters to temperature = 1.0, max\_tokens = 1000, with all other parameters set to default values. The sycophantic LLM was instructed using a single system prompt. As LLMs are biased in the sycophantic direction, the neutral LLM used a two-stage pipeline: a system prompt with no stance instructions was used to generate an initial response, and a second LLM call was then used to remove any residual validating or affirming language~\cite{zhao2025knoll}. The challenging LLM (Studies 4 and 5 only) used the same system prompt as the neutral condition along with a hidden injected user-assistant exchange where the user instructs the model to point out reasoning flaws and contradictions~\cite{sharma2023towards}. We confirmed that models behaved as intended in three ways: through pre-study benchmarking, LLM-judge evaluation of sampled experiment conversations, and participant-reported manipulation checks (see Supplementary Information, Section 2.1)~\cite{cheng2026elephant}. All experiments were conducted between January 2026 and April 2026. 

\subsection*{Study 1. Importance of support from humans and AI}
We recruited 253 U.S.-based adults via the crowd-sourcing platform Prolific (gender-balanced through Prolific pre-screening; $N = 228$ after excluding 25 participants for failed attention checks). In a between-subjects design, participants described a personal situation they were seeking advice on and were randomly assigned to rate the importance of receiving different kinds of support from a conversation with either someone close to them or an AI chatbot. Participants in the human condition identified a specific person they would normally talk to about that situation and reported the relationship type (e.g., friend, family member) and closeness using the Inclusion of Other in the Self Scale~\cite{aron1992inclusion}. Participants in the AI condition reported their AI usage frequency. All participants rated the importance of four types of support---emotional, esteem, informational, and certainty---on 7-point scales~\cite{cutrona1992controllability}. Our primary preregistered analysis tested the source $\times$ support type interaction using ANOVA, followed by simple effects and planned contrasts comparing the human-AI gap for emotional and esteem support to that for informational support.

\subsection*{Study 2. Perceptions of support in human-sycophantic AI conversations}
We recruited 400 U.S.-based adults via Prolific (gender-balanced through Prolific pre-screening; $N = 391$ after excluding nine participants for failed attention checks). Participants described a personal situation they were seeking advice on, were randomly assigned to either a sycophantic or neutral AI condition, and had an open-ended conversation with their assigned AI system about their situation. After the conversation, participants rated the AI systems on the same four support dimensions from Study 1 (emotional, esteem, informational, and certainty). Our primary preregistered analyses tested whether the sycophantic AI received higher ratings of emotional and esteem support than the neutral AI, while our exploratory preregistered analyses tested condition differences in informational support and certainty.

\subsection*{Study 3. Anticipation of human conversations following sycophantic AI}
We recruited 600 U.S.-based adults via Prolific (gender-balanced through Prolific pre-screening; $N = 592$ after excluding 6 participants for failed attention checks). Participants described a personal situation they were seeking advice on, identified a specific person they would normally go to for advice, and reported the relationship type and closeness using the Inclusion of Other in the Self Scale~\cite{aron1992inclusion}. They were randomly assigned to either a sycophantic or neutral AI condition and had an open-ended conversation with the AI systems about their situation. After the conversation, participants rated their expectations for a subsequent conversation with the identified person about the same situation, including the anticipated effort it would take to be understood by that person and the perceived sufficiency of their AI system conversation. Our primary preregistered analyses tested whether sycophantic AI increases anticipated effort and perceived sufficiency, controlling for closeness to the identified person. Exploratory preregistered analyses tested moderation by closeness and mediation by AI conversation perceptions (helpfulness, certainty, ease, and feeling understood by the AI).

\subsection*{Study 4. Longitudinal study of sycophantic AI's impact}
We recruited 1,400 U.S.-based adults via Prolific (census-representative on age, gender, and ethnicity through Prolific pre-screening) and randomly assigned them to one of four conditions: sycophantic AI, challenging AI, neutral AI, or no-AI control. Participants in AI conditions completed 12 sessions over three weeks (four per week), each consisting of open-ended conversation about a self-selected personal advice topic drawn from a pool of 16 prompts (full list in Supplementary Methods, Section 1.3). These prompts were intentionally designed around situations where the user is considering or has taken an action of questionable wisdom, either toward others (e.g., ending a friendship, refusing to help) or toward themselves (e.g., neglecting sleep, unhealthy coping habits). This created contexts where uncritical validation could carry a real cost, and where sycophancy is therefore most likely to differ from neutral or challenging responses. Sessions required a minimum of 7 conversational turns and ended after a maximum of 20. Participants in the no-AI control condition completed only the initial and final surveys and received no between-survey task. Pre- and post-treatment surveys measured three primary outcomes: feeling understood by humans, intellectual humility, and self-enhancement. A fourth primary outcome, relative preference for AI over humans for personal advice, was measured weekly. Other session-level and weekly surveys complemented these outcomes by measuring participants’ experience with the AI (e.g., helpfulness, positive affect, feeling understood by AI), social time spent with others, social satisfaction, and actions taken based on advice (see Supplementary Information, Section 2.5 for full list of secondary and exploratory outcomes). Repeated measures were analyzed with mixed-effects regression (condition, time, condition $\times$ time, random intercepts); pre-post outcomes were analyzed by regressing post-treatment scores on condition and baseline scores. Pairwise contrasts against the sycophantic condition were corrected using Holm-Bonferroni within each hypothesis family of contrasts.\footnote{Deviations from the pre-registered analysis plan: (i) the pre-registration included a global clause stating that all primary models include baseline scores as covariates. Here, we did not include the baseline relative-preference item as a covariate in the H1 mixed-effects model because that item asked about ``an AI chatbot'' in general, whereas the weekly follow-up measures asked about the participant's specific assigned chatbot, and the two are therefore not on the same construct. A sensitivity analysis that nonetheless includes the baseline item as a covariate results in an equivalent conclusion (Supplementary Information, Section 2.5). (ii) Mixed-effects models specified random intercepts and random slopes for time in the pre-registration, with the pre-registration permitting a fallback to intercepts only if convergence issues arose. All weekly models reported here use random intercepts only. (iii) Effect sizes are reported as Cohen's d using two conventions matched to the analysis type (no convention was specified in the pre-registration). For mixed-effects analyses on longitudinal outcomes (H1 and weekly trajectories), d is the fixed-effect condition coefficient divided by the between-participant standard deviation of person-level outcome means within the two contrast arms. For between-subjects analyses with one observation per participant (H2--H7 ANCOVA, manipulation checks, exploratory contrasts), d uses the pooled within-condition standard deviation of the outcome.} Attrition was 15.7\% in AI conditions (84.3\% completing all 12 sessions) and 10\% in the no-AI control (90\% completing pre- and post-treatment surveys), with no significant differential attrition across conditions ($\chi^2$ = 2.05, p = 0.36). Full attrition analyses are reported in Supplementary Information, Section 2.5.2. Missing data was handled with full-information maximum likelihood (FIML) under missing-at-random (MAR) assumptions, following intention-to-treat principles.

\subsection*{Study 5. Preferences for different AI interaction styles}
We recruited $N = 500$ U.S.-based adults via Prolific (gender-balanced through Prolific pre-screening) for a within-subjects study. Participants described a personal situation they were seeking advice on, then had brief conversations (three conversational turns each) with three AI interaction styles (sycophantic, neutral, challenging), presented as Models A, B, and C with randomized label-to-style mapping. After sampling all three, participants selected which model they would most want to continue talking to. Our primary preregistered analysis predicted that the sycophantic AI would be chosen as first preference more often than chance. Our exploratory preregistered analyses examined preference rankings, sampling order effects, and reasons for participants' choice.

\subsection*{Participants and procedures}
All studies were approved by the Stanford University Institutional Review Board. Participants provided informed consent before beginning each study. Participants were recruited from Prolific and compensated at a median rate of \$12/hour. Participants in the longitudinal study received a completion bonus of \$10 (distributed throughout the three-week period) for completing all 12 sessions. Preregistrations for all five studies are available at \url{https://osf.io/5ef7b}. Data and analysis code are available at \url{https://github.com/lujainibrahim/syc-ai-impact}. Sample sizes were determined by power analyses targeting 80-90\% power to detect effects at $\alpha$ = 0.05 (see Supplementary Information for each study’s power analysis). Across all studies, participants were excluded for failing the attention check or completing the study ``inhumanly quickly" as defined and detected by Prolific’s automatic checker. Condition assignment was implemented via Qualtrics randomizer, and participants were blind to condition. Complete item wordings, scale details, and full analysis plans for all studies are provided in Supplementary Information and the pre-registrations.

\putbib[sn-bibliography] 
\end{bibunit}

\bmhead{Author contributions}
L.I.: Conceptualization, Methodology, Software, Formal analysis, Investigation, Writing - Original Draft. F.H.: Conceptualization, Methodology, Software, Formal analysis, Investigation, Data Visualization, Writing - Review \& Editing. M.C.: Conceptualization, Investigation, Writing - Review \& Editing. C.L.: Investigation, Writing - Review \& Editing. R.A.: Project administration, Writing – Review \& Editing. R.W.: Investigation, Writing – Review \& Editing. L.R.: Conceptualization, Methodology, Supervision, Writing - Review \& Editing. D.Y.: Conceptualization, Methodology, Supervision, Writing - Review \& Editing.

\bmhead{Acknowledgments}
L.I. acknowledges funding from the UK AI Security Institute. F.H.'s PhD is supported by the Economic and Social Research Council [ES{\textbackslash}Y001761{\textbackslash}1]. M.C.'s PhD is supported by the Stanford Knight-Hennessy scholarship. L.R. acknowledges support from the Royal Society Research Grant RG{\textbackslash}R2{\textbackslash}232035 and the UKRI Future Leaders Fellowship [MR/Y015711/1]. D.Y is supported by grants from Open Philanthropy, ONR N000142412532, NSF IIS 2247357, and Schmidt Sciences. We thank Alia El Kattan and Meryl Ye for feedback on the manuscript; Andrew Strait and Julianne Ancey for support facilitating the project grant; Sunny Yu, Dora Zhao, Steve Rathje, Desmond Ong, and Verena Rieser for helpful conversations. 

\bmhead{Competing interests}
C.L. is currently employed by Microsoft. L.I. was recently in a contractual agreement with Google DeepMind. No competing financial interests exist related to the presented results. The other authors declare no competing interests.
 
\bmhead{Supplementary Information}
Supplementary Information is available at \url{https://github.com/lujainibrahim/syco-long-study}. 

\bmhead{Code availability}
Code for statistical analyses and figure generation is available at \url{https://github.com/lujainibrahim/syco-long-study}. 

\bmhead{Data availability}
Anonymized data from all experiments is available at \url{https://github.com/lujainibrahim/syco-long-study}. 

\backmatter

\noindent

\clearpage

\end{document}